\documentclass[a4paper,fleqn]{cas-dc}

\usepackage[numbers]{natbib}
\usepackage{subfig}
\usepackage{amsmath}
\usepackage{lineno}  
\usepackage{siunitx} 
\usepackage{multicol}
\usepackage{multirow}

\begin{document}
\let\WriteBookmarks\relax
\def\floatpagepagefraction{1}
\def\textpagefraction{.001}
\shorttitle{Patient Bed Horizontal Movement Performance Estimation}
\shortauthors{J. Kwak~\textit{et~al}.}

\title[mode = title]{Estimating horizontal movement performance of patient beds and the impact on emergency evacuation time}         

%
%
\author[1]{Jaeyoung Kwak}
\cormark[1]
\address[1]{Complexity Institute, Nanyang Technological University, 61 Nanyang Drive, Singapore 637335, Singapore}

\author[2]{Michael H. Lees}
\address[2]{Informatics Institute, University of Amsterdam, Science Park 904, Amsterdam 1098XH, The Netherlands}

\author[3]{Wentong Cai}
\address[3]{School of Computer Science and Engineering, Nanyang Technological University, 50 Nanyang Avenue, Singapore 639798, Singapore}

\author[4, 6, 1]{Ahmad Reza Pourghaderi}
\address[4]{Health Systems Research Center (HSRC), Singapore Health Services, 31 Third Hospital Avenue, Singapore 168753, Singapore}

\author[5, 6]{Marcus E.H. Ong}
\address[5]{Department of Emergency Medicine, Singapore General Hospital, Outram Road, Singapore 169608, Singapore}
\address[6]{Health Services and Systems Research (HSSR), Duke-NUS Medical School, 8 College Road, Singapore 169857, Singapore}

%

\begin{abstract}
Emergency evacuation of patients from a hospital can be challenging in the event of a fire. Most emergency evacuation studies are based on the assumption that pedestrians are ambulant and can egress by themselves. However, this is often not the case during emergency evacuations in healthcare facilities such as hospitals and nursing homes. To investigate emergency evacuations in such healthcare facilities, we performed a series of controlled experiments to study the dynamics of patient beds in horizontal movement. We considered a patient bed because it is one of the commonly used devices to transport patients within healthcare facilities. Through a series of controlled experiments, we examined the change of velocity in corner turning movements and speed reductions in multiple trips between both ends of a straight corridor. Based on the experimental results, we then developed a mathematical model of total evacuation time prediction for a patient bed horizontally moving in a healthcare facility. Factoring uncertainty in the horizontal movement, we produced the probability distribution of movement duration and estimated the probability that an evacuation can be safely performed within certain amount of time. In addition, we predicted that the evacuation time would be longer than the prediction results from an existing model which assumes constant movement speed. Our results from the model demonstrated good agreement with our experimental results.
\end{abstract}

%
%
%
\maketitle


\section{Introduction}
\label{intro}
Pedestrian emergency evacuation has been one of the central topics in the field of fire safety engineering. In case of life threatening incidents such as fire and hazardous chemical spills, a well-prepared emergency evacuation plan can efficiently move occupants to the assembly points with minimum amount of time and ensure the safety of evacuees. Based on the degree of required aid during egress, occupants can be categorized into ambulant and non-ambulant occupants. Ambulant occupants can egress to the place of safety without help from other occupants. On the other hand, non-ambulant occupants need help when they move, especially in case of an emergency. The non-ambulant occupants can be further categorized into subcategories such as bedridden occupants and wheelchair users depending on the device that they are using. 

Emergency evacuation of ambulant pedestrians has been investigated in experimental studies to understand pedestrian movement during emergency evacuations. Numerous experiments have been conducted in horizontal evacuation scenarios in which pedestrians egress within the same floor. For instance, researchers have studied evacuations through bottlenecks to understand the relationship between the bottleneck width and pedestrian flow, which is critical to the total evacuation time in room evacuations~\cite{Kretz2006a, Hoogendoorn_TrSci2005, Seyfried_TrSci2009}. A considerable number of experiments have been performed to characterize pedestrian vertical movement through stairs such as downward moving speed and flow rate in a stairwell of a high-rise building~\cite{Huo_SafetySci2016} and the influence of stair slope on such pedestrian flow characteristics~\cite{Burghardt_TrC2013}. The pedestrian flow characteristics have been studied for harsh moving conditions including crawling in a room~\cite{Nagai_PhysicaA2006} and in a corridor~\cite{Kady_FSJ2009}, and the existence of earthquake-induced falling debris~\cite{Lu_SafetySci2019}.

In most existing emergency evacuation studies, it is assumed that pedestrians are able to walk and egress by themselves. However, this is often not the case during emergency evacuations especially in healthcare facilities such as hospitals and nursing homes. Healthcare facilities accommodate considerable numbers of non-ambulant patients who have limited mobility and need the help of medical staff with evacuation devices during the evacuation process. To prepare emergency evacuation plans considering the non-ambulant pedestrians, it is necessary to understand the performance of evacuation devices such as their movement speed and movement dynamics. 

Previous experimental studies of non-ambulant pedestrian evacuations analyzed video footage of the experiments to measure travel time between different reference points. By doing that, average movement speed was measured for different sections of an evacuation route such as a straight corridor and a stairwell. For example, Rubadiri~\textit{et al.}~\cite{Rubadiri_FireTech1997} performed a horizontal evacuation exercise of manual and electric wheelchair users. They measured the evacuation performance index as a ratio of the wheelchair user movement speed without assistance to the movement speed of ambulant pedestrians. Based on the measured evacuation performance index, they predicted evacuation time of the wheelchair users and then compared it with the actual measurement of the wheelchair users. Strating~\cite{Strating_thesis2013} collected movement speed data of bedridden occupants in Dutch healthcare facilities. He reported an evacuation speed range from 0.54 to 1.34~m/s in Dutch hospitals and from 0.25 to 1.30~m/s in Dutch nursing homes. Hunt~\textit{et al.}~\cite{Hunt_FireMater2015} measured horizontal and vertical movement speed of evacuation devices including stretchers, evacuation chairs, carry chairs, and rescue sheets. They observed that the evacuation chair is the fastest among the tested evacuation devices with an average speed of 1.5~m/s in the horizontal evacuation and of 0.83~m/s in the vertical evacuation. Based on the measured movement speed of those evacuation devices, they also estimated the total evacuation time of non-ambulant patients in a high-rise hospital building. While those studies focused on measuring the average movement speed of non-ambulant pedestrian evacuation devices, change of velocity during the movement has not yet been studied in detail.

In pedestrian flow dynamics, it has been reported that pedestrians continuously moving with heavy load tend to have reduced movement speed due to fatigue~\cite{Luo_JSTAT2016}. Furthermore, several studies investigated speed profiles of pedestrians moving near obstacles in front of an exit~\cite{Shi_PhysicaA2019}, corners~\cite{Dias_TRR2014}, and merging areas~\cite{Shiwakoti_SafetySci2015}. Such structural elements of pedestrian facilities often act as a constraint on the efficiency of pedestrian flow seemingly because pedestrians need to change their moving direction and speed.

Similar to the case of pedestrian flow dynamics, the movement of an evacuation device becomes slower when the device is moving in corners and areas of merging flow and interaction with other devices coming from different directions. In addition, there are often far more bedridden patients than medical staff in healthcare facilities. It is apparent that, in emergency evacuations, every medical staff needs to make multiple trips between the location of bedridden patients and the place of safety. One can expect that the medical staff experience some level of exhaustion when they evacuate all the bedridden patients and it is likely that their movement becomes slower as they travel longer distance. Such a slow-moving evacuation device affects the flow of subsequent evacuation devices and pedestrians, especially with limited passing opportunities in a narrow corridor. Consequently, understanding these non-linear dynamics of evacuation devices enables us to better estimate the evacuation performance of the devices with taking into account their speed profile and change of moving directions. 

Among various evacuation devices, we considered the movement of a patient bed because it is one of the most commonly used devices to transport patients from one place to the another in healthcare facilities. In this study, we performed a series of controlled experiments to investigate the dynamics of patient beds in horizontal movement. Through the controlled experiments, we examined the change of velocity in corner turning movements and speed reductions in multiple trips between both ends of a straight corridor. Based on the experimental results, we then developed a movement duration prediction model and then applied the model for a patient bed horizontally moving in a healthcare facility. Incorporating uncertainty in the horizontal movement, we predicted probability that an evacuation can be safely performed within certain amount of time. The experiment setup and data collection methods are described in Section~\ref{setup}. As shown in Section~\ref{results}, we analyze the dynamics of patient bed movement, develop a movement duration prediction model, and then apply the model for a patient bed horizontally moving in a healthcare facility. We summarizes the results with concluding remarks in Section~\ref{conclusion}.

\section{Experiment setup}
\label{setup}
To collect patient bed movement data, we carried out a series of controlled experiments in September 2019 at the Singapore General Hospital, Singapore. We recruited 8 individuals who were acting as handlers in the experiment: 4 males and 4 females aged between 25 to 35 without movement impairments. In order to move a patient bed, the handlers were grouped into pairs of same gender handlers: male-male and female-female handlers. Additionally, one male (27 years old, 189 cm, 87 kg) was lying on the patient bed during its movement in order to substitute for a real non-ambulant patient. In total, 9 individuals (8 handlers and 1 person acting as a bedridden patient) were volunteered for the experiment. Before starting the experiment, the handlers had an orientation session and conducted a few warm-up trials under the guidance of a nurse who works at the Singapore General Hospital Emergency Department. Through the orientation session, the handlers became proficient enough in maneuvering the patient bed. In the experiment, we asked all the handlers to move the patient bed as fast as possible while making sure their safety. 

\begin{figure*}
	\centering\includegraphics[width=17.5cm]{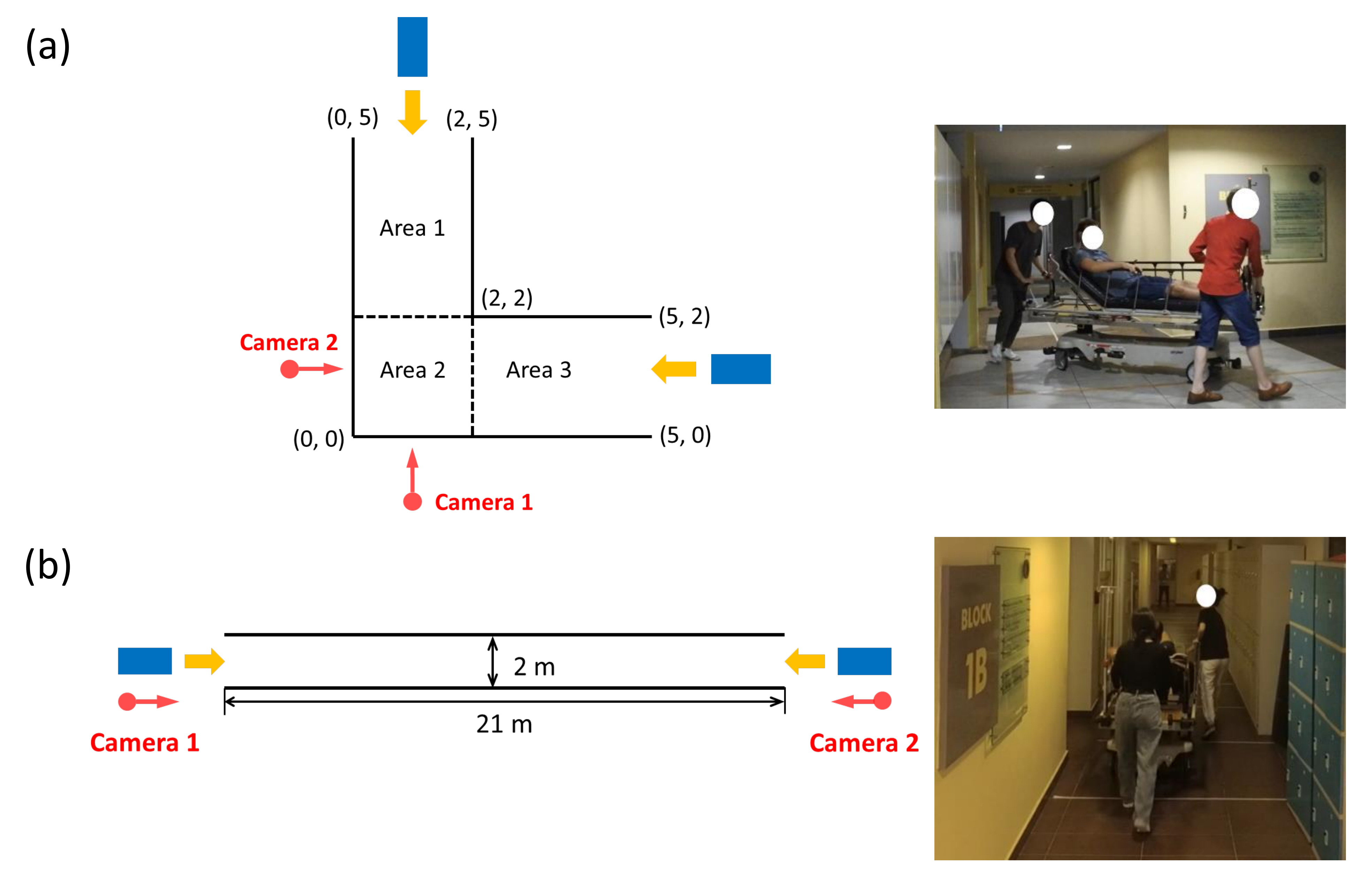}\vspace{-0.3cm}
	\caption{Schematic representation of experiment setup: (a) a \ang{90}~angled corner and (b) a straight corridor. In (a), Areas 1 and 3 indicate 3~m long 2~m wide straight corridor sections and Area 2 shows 2~m-by-2~m intersection. Blue rectangles depict patient beds and yellow arrows show their moving direction. Red circles and red arrows indicate the locations of cameras and their pointing direction, respectively.} 
	\label{fig:ExperimentSetups} 
	\vspace{-0.5cm}
\end{figure*}

\begin{figure*}
	\centering\includegraphics[width=14.5cm]{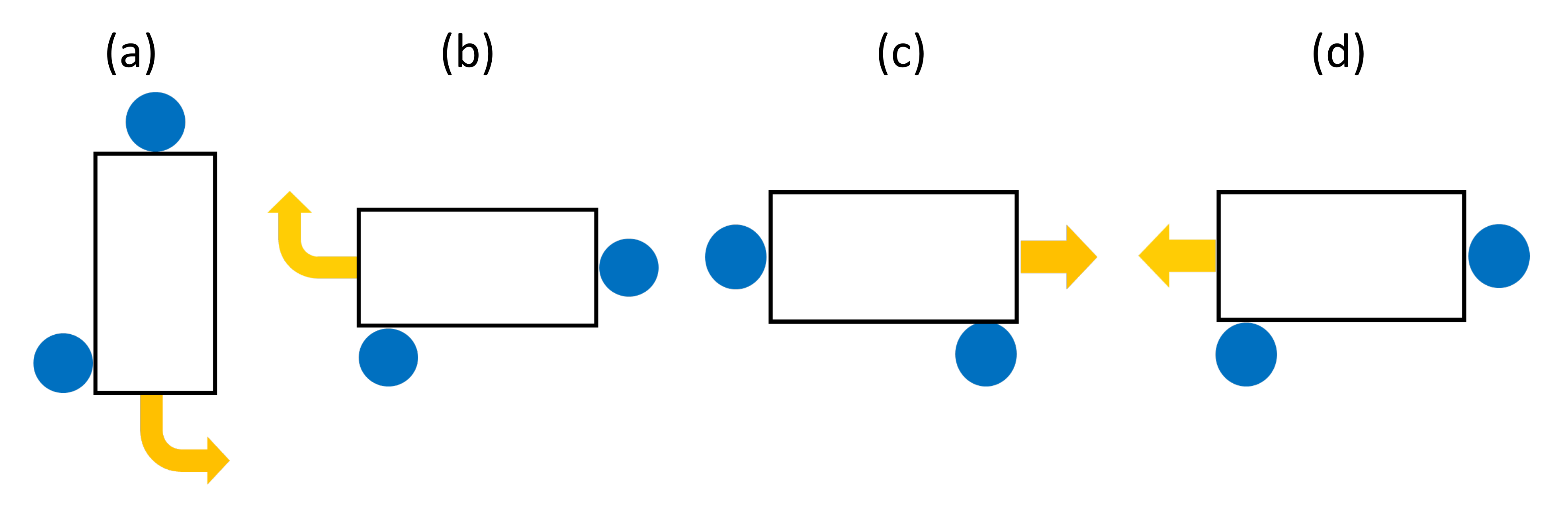}\vspace{-0.3cm}
	\caption{Schematic representation of handler positions with the direction of travel: (a) 
	left-turning movement from the top to the right branches in the \ang{90}~angled corner setup, (b) right-turning movement from the right to the top branches in the \ang{90}~angled corner setup, (c) rightward movement in the straight corridor setup, and (d) leftward movement in the straight corridor setup. Blue circles depict handlers and yellow arrows show their moving direction.} 
	\label{fig:HandlerPosition} 
	\vspace{-0.5cm}
\end{figure*}

Figure~\ref{fig:ExperimentSetups} shows the sketches of the experiment setup: a \ang{90}~angled corner and a straight corridor. In the \ang{90}~angled corner setup, each handler pair was maneuvering a patient bed in the intersection of 2~m $\times$ 2~m and two straight corridors of 2~m wide 5~m long. Handlers made left-turning movements by moving from the top to the right branches, and from the right to the top branches for right-turning movements. The handler pair repeated right and left turning movement at least 10 times. In the straight corridor experiment setup, each handler pair made 21 round trips with a patient bed in a straight corridor of 2~m wide 21~m long. By doing that, the handler pair moved the patient bed a distance of 882~m in effect. In both experiment setups, there were not obstacles or doors on the experimental route. According to the usual practice~\cite{ABHB_2012}, a pair of handlers was moving the patient bed together in both experiments. In each pair, one handler was acting as a leading handler who was guiding the movement direction of the bed, while the other handler was pushing the bed following the leading handler. Once the handlers reached the end of the experiment area, they changed their position in order to reverse their movement direction. The formation of handlers in each experiment setup is illustrated in Fig.~\ref{fig:HandlerPosition}. 

Both experiments were recorded using cameras with a frame rate of 30 frames per second. The cameras were mounted on tripods which were set on the top of tables. In the \ang{90}~angled corner setup, we set two cameras near the corner to closely observe the turning movement of the patient bed. In the straight corridor setup, two cameras were used to record the bed movement from each end of the corridor. From video footage of the \ang{90}~angled corner setup, we extracted patient bed movement trajectories using T-analyst, a semi-automatic trajectory extraction software developed from Lund University, Sweden~\cite{T-Analyst}. The software has been applied in cyclist traffic safety analysis~\cite{Fyhri_AAP2016} and pedestrian flow analysis~\cite{Nielsen_PED2014}. The conversion from video coordinates to the real world coordinates was performed by T-calibration, a calibration software accompanied by T-analyst. In T-calibration software, we placed calibration points and its real-world coordinates and then performed TSAI-calibration algorithm~\cite{Tsai_1987}. After the calibration process, we manually annotated pedestrian positions for every 10 frames on average in the video footage with T-analyst software. Next, T-analyst software then extracted pedestrian trajectories. Details of pedestrian trajectory extraction process can be found from T-analyst manual available from its webpage~\cite{T-Analyst}.

\begin{table*}
	\normalsize                       %
	\setlength{\tabcolsep}{6pt}       
	\renewcommand{\arraystretch}{1.2} 
	\centering
	\caption{Basic movement characteristics of handler groups in the corner area of $x \le 5$ and $y \le 5$.}
	\label{table-handler-1}
	\resizebox{17.0cm}{!}{
		\begin{tabular}{l*{10}{c}}
			\hline\hline 
			Handler & \multirow{2}{*}{Gender} & Movement & \multicolumn{2}{c}{Orientation (rad)} & No. of & Distance & Duration & Average & Entering \\
			\cline{4-5}
			groups  &       & direction & starting & target & trajectories & (m) & (s) & speed (m/s) & speed (m/s)\\		
			\hline
			LT\_1 & Female	& Left turn  & 1.57 & 0		& 12 & 7.59 $\pm$ 0.16 & 10.59 $\pm$ 0.57 & 0.72 $\pm$ 0.03 & 0.80 $\pm$ 0.06\\
			LT\_2 & Female	& Left turn  & 1.57 & 0		& 13 & 7.45 $\pm$ 0.11 &  9.05 $\pm$ 0.67 & 0.83 $\pm$ 0.06 & 0.87 $\pm$ 0.09\\
			LT\_3 & Male	& Left turn  & 1.57 & 0		& 10 & 7.58 $\pm$ 0.12 & 10.17 $\pm$ 0.73 & 0.75 $\pm$ 0.05 & 0.72 $\pm$ 0.15\\
			LT\_4 & Male	& Left turn  & 1.57 & 0		& 15 & 7.59 $\pm$ 0.10 &  7.34 $\pm$ 1.04 & 1.05 $\pm$ 0.11 & 0.95 $\pm$ 0.12\\
			RT\_1 & Female	& Right turn & 3.14 & 1.57	& 12 & 7.67 $\pm$ 0.06 &  9.13 $\pm$ 0.54 & 0.84 $\pm$ 0.05 & 1.15 $\pm$ 0.07\\
			RT\_2 & Female	& Right turn & 3.14 & 1.57	& 13 & 7.59 $\pm$ 0.10 &  7.24 $\pm$ 0.42 & 1.05 $\pm$ 0.06 & 1.32 $\pm$ 0.11\\
			RT\_3 & Male	& Right turn & 3.14 & 1.57	& 10 & 7.98 $\pm$ 0.12 &  9.33 $\pm$ 0.61 & 0.86 $\pm$ 0.04 & 1.05 $\pm$ 0.08\\
			RT\_4 & Male	& Right turn & 3.14 & 1.57	& 15 & 7.82 $\pm$ 0.13 &  6.47 $\pm$ 0.56 & 1.22 $\pm$ 0.08 & 1.51 $\pm$ 0.24\\
			\hline
			Average   & & & & & & 7.65 $\pm$ 0.19 & 8.51 $\pm$ 1.57 & 0.93 $\pm$ 0.18 & 1.07 $\pm$ 0.29\\
			\hline\hline
		\end{tabular}
	}
	\vspace{-0.5cm}
\end{table*}

Table~\ref{table-handler-1} shows the basic movement characteristics of handler groups in the turning movement experiment. We collected 10 to 15 trajectories from each handler group and in total $N = 100$ trajectories were collected. The distance traveled and the movement duration were measured for each trajectory in the area of $x\le5$ and $y\le5$ and then averaged for each handler group. The average speed was computed by dividing the distance traveled by the movement duration. The entering speed was measured when the patient bed is entering the area of $x\le5$ and $y\le5$. Handler groups RT\_2 and RT\_4 showed shorter movement duration than other handler groups seemingly because they entered the corner with higher entering speed and experienced smaller speed drop in the course of their movement.

\begin{table}[]
	\normalsize                 %
	\setlength{\tabcolsep}{8pt} %
	\centering
	\caption{Basic movement characteristics of handler groups in straight corridor experiment.}
	\label{table-handler-2}
	\resizebox{8.0cm}{!}{
		\begin{tabular}{ccccc}
			\hline\hline 
			Handler & \multirow{2}{*}{Gender} & \multicolumn{3}{c}{speed (m/s)}\\
			\cline{3-5}
			groups  & & min. & max. & average\\
			\hline
			Group\_1 & Female	& 0.94 & 1.20 & 1.03 \\
			Group\_2 & Female	& 1.34 & 1.64 & 1.47 \\
			Group\_3 & Male		& 0.98 & 1.13 & 1.04 \\
			Group\_4 & Male		& 1.31 & 1.70 & 1.53 \\
			\hline\hline
		\end{tabular}	
	}
	\vspace{-0.5cm}
\end{table}

In the straight corridor setup, based on the study of Luo~\textit{et al.}~\cite{Luo_JSTAT2016}, we assumed that the fatigue experienced by the handlers is affected by the distance that they transported the patient bed. The handlers were making multiple trips between one end to the other end of the corridor. We measured the time that the handlers took moving between the two ends of the corridor using the video footage. In doing that, we manually marked the transition period in which the handlers stopped and changed their position. After identifying the transition period, we obtained the travel time in each one-way trip and then calculated the average speed for each trip. The initial speed was measured from the first one-way trip in which the handlers moved first 21 meters in the straight corridor setup. Our approach of measuring the speed is analogous to the method presented by Chen~\textit{et al.}~\cite{Chen_FireTech2017}. The average speed was measured by dividing the distance traveled (882 m) by the total travel time that the handlers moving the patient bed. Table~\ref{table-handler-2} summarizes the basic movement characteristics of handler group in the straight corridor experiment. Handler groups Group\_2 and Group\_4 showed higher speed but there is not significant difference between male and female. The differences between the groups in the same gender (Group\_1 vs. Group\_2, and Group\_3 vs. Group\_4 in Table~\ref{table-handler-2}) seemingly attribute to the degree of risk aversion set by different groups. Although we asked all the handlers to move as fast as possible, some of them might want to more focus on moving safely without getting injured, so they were not necessarily in a rush.

\section{Results and analysis}
\label{results}
\subsection{Turning movement}

\begin{figure}
	\centering\includegraphics[width=0.75\columnwidth]{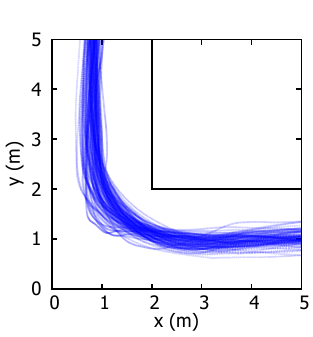}\vspace{-0.3cm}
	\caption{$N = 100$ individual trajectories collected from the corner turning experiment setup} 
	\label{fig:Trajectories} 
\end{figure}

Figure~\ref{fig:Trajectories} shows individual trajectories collected from the corner turning experiment setup. To understand the change of velocity in the course of turning movement, we firstly obtained the average path of all the handler groups moving in both movement directions. Similar to Hicheur~\textit{et al.}~\cite{Hicheur_EJN2007}, we resampled the collected trajectories into $N_f = 60$ equal intervals and then averaged the resampled trajectories for each rescaled time point $\hat{t} \in [0, 1]$:
\begin{equation}
\label{eq:AvgPath} 
\begin{split}
x_{avg}(\hat{t}) &=& \frac{1}{N} \sum_{i = 1}^{N} x_i(\hat{t}),\\ 
y_{avg}(\hat{t}) &=& \frac{1}{N} \sum_{i = 1}^{N} y_i(\hat{t}).
\end{split}
\end{equation}
\noindent
Here, $N = 100$ is the number of collected trajectories from handler groups shown in Table~\ref{table-handler-1}. Likewise, we calculated trajectory deviation $\sigma (\hat{t})$ to quantify the difference between the mean trajectory and an individual trajectory $i$:

\begin{equation}
	\label{eq:path_x_sd} 
	\begin{split}
	\sigma_x (\hat{t}) &=& \sqrt{\frac{1}{N} \sum_{i = 1}^{N} \left( x_i(\hat{t})-x_{avg}(\hat{t}) \right)^2},\\ 
	\end{split}
\end{equation}

\begin{equation}
	\label{eq:path_y_sd} 
	\begin{split}
	\sigma_y (\hat{t}) &=& \sqrt{\frac{1}{N} \sum_{i = 1}^{N} \left( y_i(\hat{t})-y_{avg}(\hat{t}) \right)^2 },\\ 
	\end{split}
\end{equation}

\begin{equation}
	\label{eq:PathDeviations} 
	\begin{split}
	\sigma (\hat{t}) &=& \sqrt{\sigma_x^2 (\hat{t}) + \sigma_y^2 (\hat{t})}. \\
	\end{split}
\end{equation}

By making use of the rescaled time points, we identified start and end of turning movement. We measured the angular displacement $\theta$ and its average $\theta_{avg}$: 
\begin{equation}
\begin{split}
\theta(\hat{t}_j)		&=& \arccos \left( \frac{\vec{e}_0 \vec{e}_i}{\left \| \vec{e}_0 \right \| \left \| \vec{e}_i \right \|} \right),\\
\theta_{avg}(\hat{t}_j) &=& \frac{1}{N} \sum_{j = 1}^{N} \theta(\hat{t}_j),	
\end{split}
\end{equation}
\noindent
where $\vec{e}_0$ is the initial moving direction which is set as (0, -1) for left-turn movement and (-1, 0) for right-turn movement. The moving direction at rescaled time $\hat{t}_j$ is given as $\vec{e}_i = (\Delta x/\Delta l, \Delta y/\Delta l)$. Here, $\Delta x$ and $\Delta y$ denote the difference between the values of $x_i$ and $y_i$ at the current rescaled time points $\hat{t}_j$ and the previous rescaled time point $\hat{t}_{j-1}$. The displacement between the positions at rescaled time $\hat{t}_j$ and $\hat{t}_{j-1}$ is given as $\Delta l = \sqrt{\Delta x^2 + \Delta y^2}$.

\begin{figure}
	\begin{tabular}{c}
		\includegraphics[width=0.95\columnwidth]{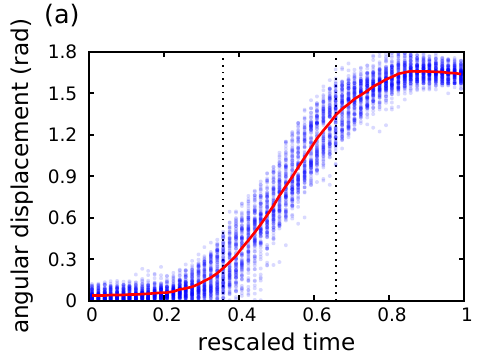}\vspace{-0.3cm}\\
		\includegraphics[width=0.95\columnwidth]{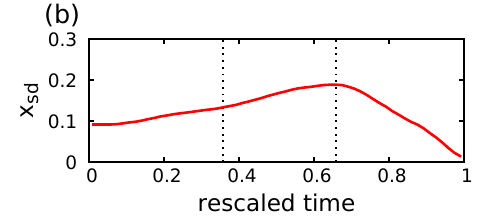}\vspace{-0.3cm}\\
		\includegraphics[width=0.95\columnwidth]{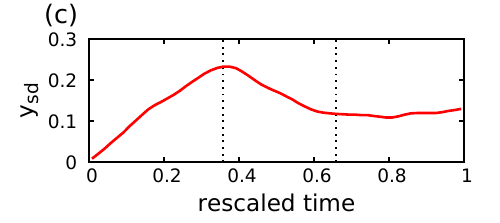}\vspace{-0.3cm}\\
		\includegraphics[width=0.95\columnwidth]{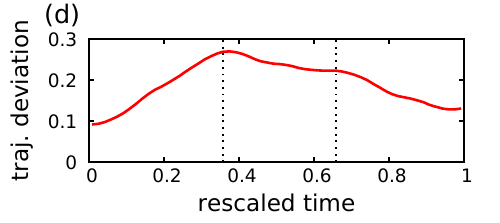}\vspace{-0.3cm}\\
	\end{tabular}	
	\caption{Angular displacement of the patient bed: (a) individual angular displacement (blue circles) and the average value (red solid line). Black dashed lines indicate start (left dashed line) and end (right dashed line) of turning movement. (b)-(d) the corresponding graphs of $\sigma_x (\hat{t})$, $\sigma_y (\hat{t})$, and $\sigma (\hat{t})$. It can be observed that the start and end of turning movement occur when either $\sigma_x (\hat{t})$ or $\sigma_y (\hat{t})$ is at peak value.} 
	\label{fig:Turning} 
\end{figure}

Figure~\ref{fig:Turning}(a) presents angular displacement of the patient bed against rescaled time points. One can observe that the angular displacement curve is nearly straight between $\hat{t} = 0.358$ and $\hat{t} = 0.658$, indicating that the patient bed is turning with a constant angular speed 0.24 rad/$\hat{t}$. When the patient bed starts and ends the turning movement, either $\sigma_x (\hat{t})$ or $\sigma_y (\hat{t})$ is at peak value. Figures~\ref{fig:Turning}(b) and~\ref{fig:Turning}(c) indicate that the start and end of turning movement can be identified based on the components of trajectory deviation, i.e., $\sigma_x (\hat{t})$ and $\sigma_y (\hat{t})$.

\begin{figure}
	\centering\includegraphics[width=0.75\columnwidth]{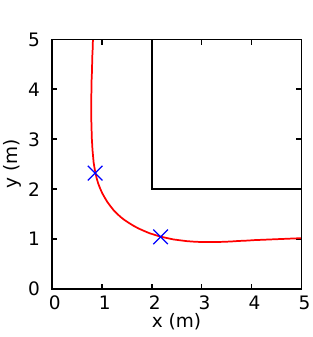}\vspace{-0.3cm}
	\caption{Average path (red solid line) with the locations of turning movement start and end (blue cross symbols).} 
	\label{fig:AveragePath} 
\end{figure}

The identified locations of turning movement start and end are presented with the mean trajectory in Fig.~\ref{fig:AveragePath}. Similar to work of Dias~\textit{et al.}~\cite{Dias_PhysLetA2018}, it was observed that the turning movement started before and ended after Area 2 ($x \le 2$ and $y \le 2$). Although there exist similarities between the patient bed and individual pedestrians in terms of the turning movement start and end locations, one can notice difference in the mean shy-away distance to the corner boundaries. Here, the mean shy-away distance was measured as the gap between the mean trajectory and the corner boundaries. In this study, the mean shy-away distance of a patient bed turning around a \ang{90}~angled corner was measured as around 1.1~m, which is larger than that one measured for individuals in the work of Dias~\textit{et al.}~\cite{Dias_Sustainable2019}. This is seemingly because the patient bed (0.8~m wide) requires larger lateral clearance than an individual (0.4 to 0.5~m wide) during the turning movement.

\begin{figure}
	\centering
	\begin{tabular}{c}
		\includegraphics[width=0.95\columnwidth]{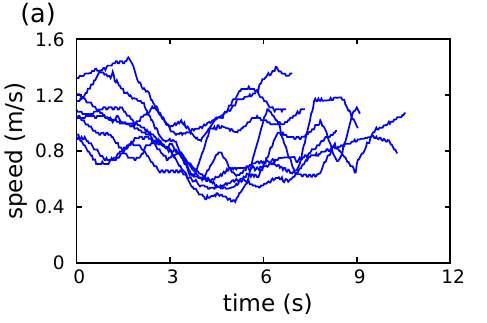}\vspace{-0.3cm}\\
		\includegraphics[width=0.95\columnwidth]{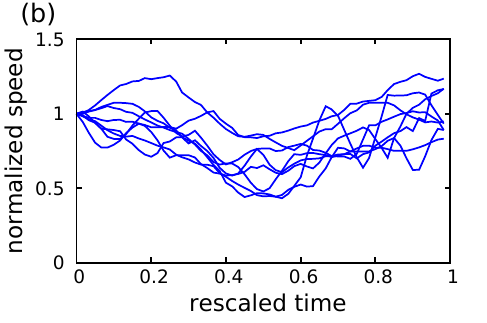}\vspace{-0.3cm}\\
		\includegraphics[width=0.95\columnwidth]{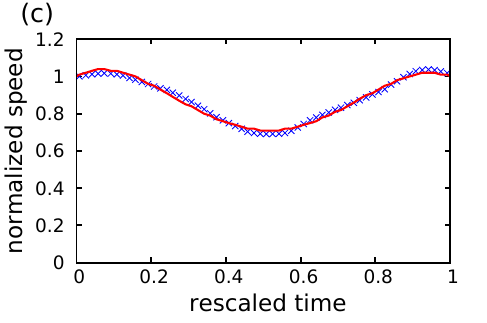}\vspace{-0.3cm}\\
	\end{tabular}
	\caption{Speed profile of handler groups as a function of rescaled time in the area of $x \le 5$ and $y \le 5$: (a) examples of individual speed curves. (b) the result of normalization performed against rescaled time $\hat{t}$ for curves in (a). (c) the average of normalized speed curves (blue cross symbols) with its trend line (red solid line). The trend line begins to decrease and then increase when the patient bed is turning around a corner.} 
	\label{fig:speed_profile} 
\end{figure}

Figure~\ref{fig:speed_profile}(a) shows examples of individual speed curves before resampling. Due to various initial speed and maneuvering duration, it is difficult to understand patterns in the speed curves. We applied the idea of the rescaled time points to such speed curves. We resampled the instantaneous speed $v_i$ of individual handler groups into $N_f$ equal intervals and then averaged the resampled speed for each rescaled time point $\hat{t}\in\left[0, 1\right]$:
\begin{equation}
\label{eq:AvgSpeed}
\begin{split}
v_{avg}(\hat{t}) &=& \frac{1}{N} \sum_{i = 1}^{N} v_i(\hat{t}).
\end{split}
\end{equation}
In order to compare different speed curves, we normalized the speed curves with entering speed $v_0$. Figure~\ref{fig:speed_profile}(b) presents the result of normalization performed against rescaled time $\hat{t}$ for curves in Fig.~\ref{fig:speed_profile}(a). We performed normalization for $N = 100$ individual speed curves and then averaged their normalized speed in rescaled time space. Figure~\ref{fig:speed_profile}(c) illustrates the average of normalized speed curves and its trend line. The trend line is generated by the normalized speed profile $\hat{v}(\hat{t})$.  

According to minimum jerk principle (Hogan~\cite{Hogan_JNeurosci1984} and Pham~\textit{et al.}~\cite{Pham_EJN2007}), people tend to minimize jerk (i.e., a third-order derivative of position) while turning around a corner:
\begin{equation}
\label{eq:MinJerkPrinciple} 
\begin{split}
\int_{0}^{1} \left[ \left( \frac{d^3 x}{d\hat{t}^3} \right)^2 + \left( \frac{d^3 y}{d\hat{t}^3} \right)^2 \right] d\hat{t},
\end{split}
\end{equation}
\noindent
where $x$ and $y$ represent position in $x$- and $y$-axis at a rescaled time point $\hat{t}$. Based on the minimum jerk principle, the normalized speed profile $\hat{v}(\hat{t})$ can be represented by a fourth-order polynomial equation of rescaled time point $\hat{t}$:
\begin{equation}
\label{eq:v_hat} 
\begin{split}
\hat{v}(\hat{t}) &=& a_0+a_1\hat{t}+a_2{\hat{t}}^2+a_3{\hat{t}}^3+a_4{\hat{t}}^4.
\end{split}
\end{equation}
\noindent
The coefficients $a_0$, $a_1$, $a_2$, $a_3$, and $a_4$ can be determined from the average of normalized speed curves in Fig.~\ref{fig:speed_profile}(c). Table~\ref{table-coefficients} shows the coefficient values. Following the work of Dias~\textit{et al.}~\cite{Dias_Sustainable2019}, we modeled the normalized speed profile $\hat{v}(\hat{t})$ with the same values in the start and end of the curve: $\hat{v}(0) = 1$ and $\hat{v}(1) = 1$.

\begin{table}
	\normalsize                 %
	\setlength{\tabcolsep}{6pt} %
	\centering
	\caption{Normalized speed profile coefficients}
	\label{table-coefficients}
	\resizebox{4.0cm}{!}{
		\begin{tabular}{cr}
			\hline\hline 
			Coefficient & \multicolumn{1}{c}{Value} \\
			\hline
			$a_0$ & 1 \\
			$a_1$ & 1.1651 \\
			$a_2$ & -10.0789 \\
			$a_3$ & 17.4769 \\
			$a_4$ & -8.5632 \\
			\hline\hline
		\end{tabular}
	}
\end{table}

\subsection{Fatigue effect}
\label{subsec:fatigue}
\begin{figure}
	\centering
	\begin{tabular}{c}
		\includegraphics[width=0.95\columnwidth]{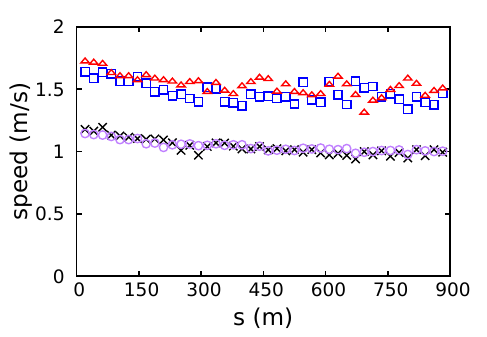}\vspace{-0.3cm}	
	\end{tabular}
	\caption{The handler group movement speed in the straight corridor setup. Different symbols represent the speed of different handler groups: black cross ($\times$) for Group\_1, blue rectangle ($\square$) for Group\_2, purple circle ($\bigcirc$) for Group\_3, and red triangle ($\bigtriangleup$) for Group\_4.} 
	\label{fig:speed_straight} 
\end{figure}

As stated in the last paragraph of Section~\ref{setup}, we measured the travel time in each one-way trip and then calculated the average speed for each trip. The handler group speed in the straight corridor setup is shown in Fig.~\ref{fig:speed_straight}. As can be seen from Fig.~\ref{fig:speed_straight}, one can observe that the handler group movement speed decreases as the distance traveled $s$ increases, especially for first 300 $m$. 

\begin{figure}
	\centering
	\begin{tabular}{c}
		\includegraphics[width=0.95\columnwidth]{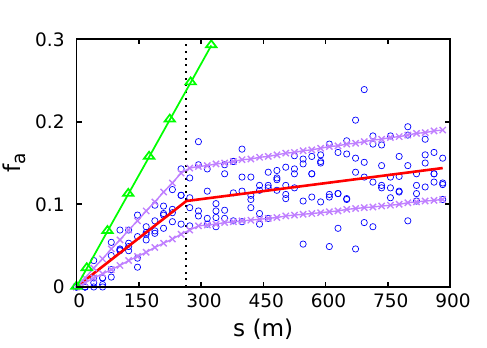}\vspace{-0.3cm}	
	\end{tabular}
	\caption{Relationship between the fatigue coefficient $f_a$ and the distance traveled $s$. The measured fatigue coefficient $f_a$ from handler groups is denoted by blue circles ($\bigcirc$). The corresponding trend line is indicated by red solid line. Black dashed line indicates the breakpoint location $s_{\times} = 264.4$~m at which fatigue coefficient $f_a$ curve slope changes. Purple solid lines with cross symbols ($\times$) show the boundaries of 80\% prediction band with quantiles 0.10 and 0.90. The fatigue coefficient curve of Luo~\textit{et al.}~\cite{Luo_JSTAT2016} is denoted by green solid line with triangles ($\bigtriangleup$). In their study, $f_a$ was measured for pedestrians carrying heavy items (10$\sim$20~kg) by hand.} 
	\label{fig:Fatigue} 
\end{figure}

To further quantify the tendency of decreasing travel speed due to fatigue effect, we evaluated the fatigue coefficient $f_a$ based on the speed reduction with respect to the initial movement speed $v_0$. According to the work of Luo~\textit{et al.}~\cite{Luo_JSTAT2016}, the fatigue coefficient $f_a$ is given as:
\begin{equation}
\label{eq:Fatigue} 
\begin{split}
f_a &=& \frac{v_0-v_i}{v_0}.
\end{split}
\end{equation}
\noindent
Here, $v_i$ indicates the current speed which was calculated for each one-way trip from one end to the other end of the straight corridor (see Fig.~\ref{fig:ExperimentSetups}(b)). As shown in Eq.~(\ref{eq:Fatigue}), the fatigue coefficient examines speed reduction ratio, so the fatigue effect can be compared among different handler groups even if they have different speed. In addition, the concept of fatigue coefficient is useful when one wants to estimate the speed at a certain traversed distance based on the initial speed. Based on Eq.~(\ref{eq:Fatigue}), we evaluated the fatigue coefficient $f_a$ for each handler group indicated in Table~\ref{table-handler-2}. Note that the current speed $v_i$ and fatigue coefficient $f_a$ are given as functions of distance traveled $s$, and $v_{i}$ is the average speed calculated from each one-way trip. Figure~\ref{fig:Fatigue} shows the relationship between $f_a$ and $s$. One can observe that the fatigue coefficient values tend to rapidly increase in the beginning and then slowly grow. That is compatible with the decreasing trend of handler group movement speed shown in Fig.~\ref{fig:speed_straight}. 

We applied piecewise linear regression~\cite{Hocking_2013} to quantify those two regimes in the fatigue coefficient curve:
\begin{equation}
\label{eq:PiecewiseLinear} 
f_a(s) = 
\begin{cases} 
\beta_{1}s 					& \text{if } s \leq s_{\times} \\
\beta_{1}s+\beta_{2}(s-s_x)	& \text{if } s > s_{\times} \\
\end{cases}
\end{equation}
where $s_{\times}$ is the breakpoint and $\beta_1$ and $\beta_2$ are regression coefficients. We used \textit{nls} function in R package to fit the piecewise linear regression model in Eq.~(\ref{eq:PiecewiseLinear}). It was estimated that the fatigue coefficient $f_a$ curve slope changes at breakpoint $s_{\times}=264.4$~m. The slope for the first and second segments are estimated as $\beta_1 = 4.076 \times 10^{-4}$ and $\beta_1+\beta_2 = 5.84 \times 10^{-5}$, respectively. The slope coefficients $\beta_1$ and $\beta_1+\beta_2$ indicate that, the speed decreases by roughly 4.08\% and 0.58\% on average for each increase of 100~m in distance traveled for $s \leq s_{\times}$ and $s > s_{\times}$, respectively. Table~\ref{table-PLRM} depicts the fatigue coefficients estimated by piecewise linear regression including the breakpoint and slopes. In addition, we also applied \textit{quantreg} function in R package to perform piecewise linear quantile regression based on Eq.~(\ref{eq:PiecewiseLinear}) in order to generate 80\% prediction band showing possible upper and lower boundaries of data points. Table~\ref{table-PLQM} depicts the slope coefficients estimated by piecewise linear quantile regression for 80\% prediction band with quantiles 0.10 and 0.90.

\begin{table}
	\normalsize                 %
	\setlength{\tabcolsep}{8pt} %
	\centering
	\caption{The fatigue coefficients estimated by piecewise linear regression: the breakpoint and slopes.}
	\label{table-PLRM}
	\resizebox{5.0cm}{!}{
		\begin{tabular}{cr}
			\hline\hline 
			Coefficient & \multicolumn{1}{c}{Value} \\
			\hline
			$s_{\times}$ & 264.4 $\pm$ 24.7 \\ 
			$\beta_1$ & $4.076 \times 10^{-4}$ \\
			$\beta_2$ & $-3.492 \times 10^{-4}$\\ 
			$\beta_1+\beta_2$ & $5.840 \times 10^{-5}$ \\
			\hline\hline
		\end{tabular}
	}
\end{table}

\begin{table}
	\normalsize                 %
	\setlength{\tabcolsep}{8pt} %
	\centering
	\caption{The slope coefficients estimated by piecewise linear quantile regression for 80\% prediction band with quantiles 0.10 and 0.90.}
	\label{table-PLQM}
	\resizebox{7.5cm}{!}{
		\begin{tabular}{crr}
			\hline\hline 
			\multirow{2}{*}{Coefficient} & \multicolumn{2}{c}{Quantiles}\\
			\cline{2-3}
			& \multicolumn{1}{c}{0.1} & \multicolumn{1}{c}{0.9} \\
			\hline
			$\beta_1$ & $2.520 \times 10^{-4}$  & $5.476 \times 10^{-4}$ \\
			$\beta_2$ & $-1.976 \times 10^{-4}$ & $-4.714 \times 10^{-4}$ \\
			$\beta_1+\beta_2$ & $5.447 \times 10^{-5}$ & $7.619 \times 10^{-5}$ \\
			\hline\hline
		\end{tabular}
	}
\end{table}

We compared the value of fatigue coefficient $f_a$ obtained in this study with that one reported in the study of Luo~\textit{et al.}~\cite{Luo_JSTAT2016}. In their study, the handlers were walking on a circular track while carrying heavy items (10$\sim$20~kg) by hand. For the same initial speed (between 1.25~m/s and 1.75~m/s), at distance of 325~m, the value of $f_a$ measured in this study ($f_a =0.11$) is lower than the value ($f_a = 0.42$) reported by Luo~\textit{et al.}~\cite{Luo_JSTAT2016}. This is seemingly because maneuvering a patient bed is physically less demanding than carrying heavy items by hand while walking. Although the data points obtained from this study are scattered in Fig.~\ref{fig:Fatigue}, it still clearly illustrates the tendency of increasing fatigue coefficient $f_a$ as the distance $s$ increases.

\subsection{Prediction of movement duration}
\label{subsec:prediction}
Based on the average normalized speed curves shown in Fig.~\ref{fig:speed_profile}(c), we can predict the movement duration of a patient bed in the corner area (i.e., $x \le 5$ and $y \le 5$). We describe the distance traveled in the corner area $s_c$ with the following equation:
\begin{equation}
\label{eq:distance_traveled} 
\begin{split}
s_c	&=& \sum_{i = 1}^{N_t} v_i \Delta t_i\\
	&=& \Delta t_i \sum_{i = 1}^{N_t} v_i,
\end{split}
\end{equation}
\noindent
where $N_t$ is the number of intervals and $v_i$ is the speed at interval $i$. The interval length $\Delta t_i$ can be obtained by rearranging Eq.~(\ref{eq:distance_traveled}):

\begin{equation}
\label{eq:delta_t_i} 
\begin{split}
\Delta t_i &=& \frac{s_c}{\sum_{i = 1}^{N_t} v_i}.
\end{split}
\end{equation}
\noindent
The movement duration in the corner area $t_c$ is a summation of $\Delta t_i$, 
\begin{equation}
\label{eq:t_c} 
\begin{split}
t_c &=& \sum_{i = 1}^{N_t} \Delta t_i\\
	&=& N_t \Delta t_i.
\end{split}
\end{equation}
\noindent
The speed at interval $i$ can be obtained from $\hat{v}_i$ by multiplying entering speed $v_0$, i.e., $v_i=v_0{\hat{v}}_i$. Here, $\hat{v}_i$ is normalized speed profile $\hat{v}$ at interval $i$. The movement duration in the corner area $t_c$ is given as:
\begin{equation}
\label{eq:movement_duration} 
\begin{split}
t_c &=& \frac{s_c N_t}{v_0 \sum_{i = 1}^{N_t} \hat{v_i}}.
\end{split}
\end{equation}
\noindent
The average normalized speed profile $\hat{v}_{avg}$ is given as:
\begin{equation}
\label{eq:avg_normalized_speed_discrete} 
\begin{split}
\hat{v}_{avg} &=& \frac{1}{N_t} \sum_{i = 1}^{N_t} \hat{v}_i
\end{split}
\end{equation}
\noindent
and if we approximate Eq.~(\ref{eq:avg_normalized_speed_discrete}) to continuous space by utilizing Eq.~(\ref{eq:v_hat}), $\hat{v}_{avg}$ becomes
\begin{equation}
\label{eq:avg_normalized_speed_continuous}
\begin{split}
\hat{v}_{avg}	&= \int_{0}^{1} \hat{v}_i d\hat{t} \\
&= \left. a_0\hat{t}+\frac{1}{2}a_1\hat{t}^2+\frac{1}{3}a_2\hat{t}^3+\frac{1}{4}a_3\hat{t}^4
+\frac{1}{5}a_4\hat{t}^5 \right|_0^1 \\
&= \left( a_0+\frac{1}{2}a_1+\frac{1}{3}a_2+\frac{1}{4}a_3+\frac{1}{5}a_4 \right).		
\end{split}
\end{equation}
\noindent
Accordingly, Eq.~(\ref{eq:movement_duration}) should read
\begin{equation}
\label{eq:movement_duration_final} 
\begin{split}
t_c &=& \frac{s_c}{v_0 \hat{v}_{avg}}.
\end{split}
\end{equation}
\noindent
Here, distance traveled in the corner area $s_c$, entering speed $v_0$, and average normalized speed profile $\hat{v}_{avg}$ are input parameters. Based on Eq.~(\ref{eq:avg_normalized_speed_continuous}), we can obtain $\hat{v}_{avg} = 0.874$. 

In order to reflect uncertainty in the movement duration of a bedridden patient (see Eq.~(\ref{eq:movement_duration_final})), we developed a probabilistic model based on the input parameters $s_c$ and $v_0$ for movement duration prediction. Previous studies~\cite{Ronchi_FSJ2014, Zhang_SafetySci2017} noted that a probabilistic approach is beneficial to estimate evacuation time based on an empirical predictive model utilizing experimental data, especially when there exists notable variability in evacuation parameters such as premovement time and movement time. In the presented experimental results, one can notice that the difference in basic movement characteristics among handlers is considerable, thus our prediction model was developed in line with the probabilistic approach. Analogous to previous studies~\cite{Rubadiri_FireTech1997, Strating_thesis2013, Hunt_FireMater2015}, the evacuation in this study refers to the movement of a bedridden patient moving with a group of handlers to the place of safety. 

\begin{figure}
	\centering\includegraphics[width=\columnwidth]{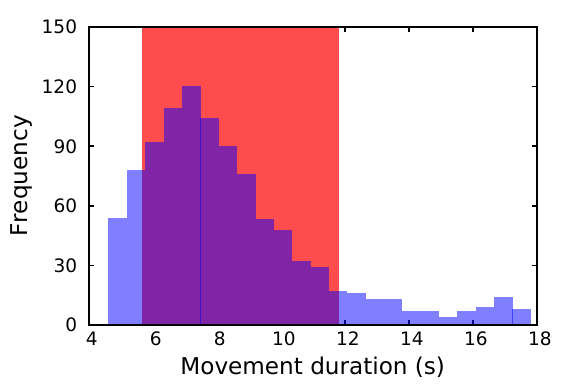}
	\caption{Frequency histogram of movement duration in a corner area estimated from numerical simulation. Red shaded area indicates the range of movement duration measured from the experiment.} 
	\label{fig:travel_time_prediction} 
\end{figure}

In the development of probabilistic prediction model, we assumed that $s_c$ and $v_0$ follow Gaussian distribution. Table~\ref{table-ProbModel_input_test} shows the probabilistic model input parameters. We set the mean and standard deviation values of the input parameters same as in the experimental results. The maximum and minimum values were determined by adding and subtracting twice of the standard deviation to and from the mean value, respectively. Based on Eq.~(\ref{eq:movement_duration_final}) with the probabilistic model input parameters in Table~\ref{table-ProbModel_input_test}, we performed 1000 simulations. The frequency histogram of the movement duration is shown in Fig.~\ref{fig:travel_time_prediction}, which has a mean of 8.85~s and a standard deviation of $2.79$~s. Its range is $[5.01, 18.62]$~s. The experimental results of movement duration $[5.67, 11.80]$~s all fall within the simulated distribution. This indicates that our prediction model of movement duration can successfully replicate the experimental data of movement duration.

\begin{table}
	\normalsize                 %
	\setlength{\tabcolsep}{8pt} %
	\centering
	\caption{Probabilistic model input parameters.}
	\label{table-ProbModel_input_test}
	\resizebox{7.5cm}{!}{
		\begin{tabular}{ccccc}
			\hline\hline 
			parameters & mean & sd & min & max\\
			\hline
			$s_c$	& 7.65 & 0.19 & 7.27 & 8.03\\
			$v_0$	& 1.07 & 0.29 & 0.49 & 1.65\\
			\hline\hline
		\end{tabular}
	}
\end{table}

In order to reflect the effect of fatigue on travel time, we considered the fatigue coefficient $f_a$ and the relationship with the distance traveled $s$ in the estimation of movement duration. Like Eq.~(\ref{eq:t_c}), the movement duration in a straight corridor $t_s$ is given as
\begin{equation}
\begin{split}
t_s &=& \sum_{i = 1}^{N_s} \Delta t_j,
\end{split}
\end{equation}
\noindent
where $\Delta t_j$ is the travel time in segment $j$ and $N_s$ is the number of segments. Note that $\Delta t_j$ is not constant here, meaning that the travel speed is different for each segment. The normalized speed $\hat{v}$ is obtained after we rearranged Eq.~(\ref{eq:Fatigue}),
\begin{equation}
\label{eq:NormalizedSpeed_Fatigue} 
\begin{split}
\hat{v} &=& \frac{v}{v_0} = 1-f_a,
\end{split}
\end{equation}
\noindent
again, $v_0$ is entering speed. The average travel speed in segment $j$ is given as
\begin{equation}
\label{eq:TravelSpeed_segment_i} 
\begin{split}
\frac{\Delta s_j}{\Delta t_j} &=& \frac{v_{j-1}+v_{j}}{2},
\end{split}
\end{equation}
\noindent
where $\Delta s_j$ is the distance traveled in segment $j$ and $v_{j-1}$ is the speed at the beginning point of segment $j$ and $v_{j}$ for the end point of segment $j$. After combining Eqs.~(\ref{eq:NormalizedSpeed_Fatigue}) and~(\ref{eq:TravelSpeed_segment_i}), we obtained the travel time of segment $j$ as
\begin{equation}
\begin{split}
\label{eq:TravelTime_segment_j} 
\Delta t_j &=& \frac{\Delta s_j}{v_0} \left( \frac{2}{\hat{v}_{j-1}+\hat{v}_{j}} \right).
\end{split}
\end{equation}

\subsection{Case study: Singapore General Hospital Emergency Department}
\label{subsec:case_study}
\begin{figure}
	\centering\includegraphics[width=0.85\columnwidth]{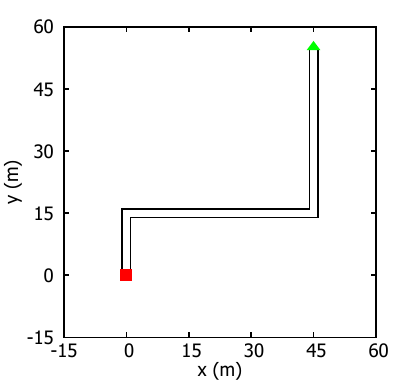}\vspace{-0.3cm}
	\caption{Schematic representation of bedridden patient evacuation route in current Singapore General Hospital Emergency Department (SGH ED). The evacuation route starts from the critical care unit (red square) which is placed at $(0,0)$ to the place of safety (green triangle) at $(45, 55)$. The evacuation route includes three straight corridor sections and two corner areas. The evacuation route is 2~m wide, same as the experiment condition.} 
	\label{fig:EvacRoute_SGH_ED} 
\end{figure}

\begin{table}
	\normalsize                 %
	\setlength{\tabcolsep}{4pt} %
	\centering
	\caption{Bedridden patient evacuation route segments}
	\label{table-EvacRouteSegments}
	\resizebox{8.5cm}{!}{
		\begin{tabular}{ccccc}
			\hline\hline 
			& Segment type & from & to & Length (m)\\
			\hline
			1 & Straight corridor	& (0, 0)   & (0, 11)  & 11 \\
			2 & Corner				& (0, 11)  & (4, 15)  & 7.65 $\pm$ 0.19 \\
			3 & Straight corridor	& (4, 15)  & (41, 15) & 37 \\
			4 & Corner				& (41, 15) & (45, 19) & 7.65 $\pm$ 0.19 \\
			5 & Straight corridor	& (45, 19) & (45, 55) & 36 \\
			\hline\hline						
		\end{tabular}
	}
\end{table}

In this case study, we demonstrate how the presented approach can be applied to predict the movement duration of a handler group transporting multiple bedridden patients. We selected the current Singapore General Hospital Emergency Department (SGH ED) for the case study. In Fig.~\ref{fig:EvacRoute_SGH_ED}, we present a schematic representation of the bedridden patient evacuation route in the current SGH ED. The evacuation route starts from the critical care unit (CCU) at $(0,0)$ to the place of safety at $(45, 55)$. The evacuation route is 2~m wide (the same width as our experiment condition), and includes three straight corridor sections and two corner areas, as shown in Table~\ref{table-EvacRouteSegments}. Note that we modeled CCU and the place of safety as single points in order to focus on the evacuation of bedridden patients between those areas. The actual geometry of the areas is much more complicated, thus the movement duration within the areas is not considered for simplicity. As in Section~\ref{subsec:prediction}, the evacuation time in this case study was estimated for a patient bed moving with two handlers. Another important point to note is that evacuation of a bedridden patient with a handler group cannot be simply extended to the case of a population of bedridden patients with multiple handler groups.This is mainly because the interactions among the handler groups are likely to affect their movement speed.

We estimated the movement duration of a handler group carrying a bedridden patient from CCU to the place of safety five times. It is reasonable to suppose that the handler group in this case study were expected to move the patient bed as fast as possible and their movement characteristics are the same as in the experimental study. Equations~(\ref{eq:movement_duration_final}) and~(\ref{eq:TravelTime_segment_j}) were applied to estimate the movement duration in the corner areas and straight corridor sections, respectively. We assumed that the fatigue effect is in effect both in the outgoing and return trips to the CCU, thus we applied the same fatigue coefficient function in Eq.~(\ref{eq:PiecewiseLinear}) for the both trips. 

\begin{table*}
	\normalsize                 %
	\setlength{\tabcolsep}{4pt} %
	\centering
	\caption{Probabilistic model input parameters.}
	\label{table-ProbModel_input_case}
	\resizebox{12cm}{!}{
		\begin{tabular}{ccccc}
			\hline\hline 
			parameters & mean & sd & min & max\\
			\hline
			$s_c$	& 7.65  & 0.19 & 7.27  & 8.03 \\
			$v_{0}$	& 1.07  & 0.29 & 0.49  & 1.65 \\
			$t_{p}$ & 7.96  & 5.19 & 4.10  & 17.64 \\
			$t_{q}$ & 6.25  & 2.60 & 2.54  & 10.25 \\			
			$s_{\times}$ & 264.4 & 24.7 & 215.0 & 313.8 \\
			$\beta_{1}$ & $4.076 \times 10^{-4}$ & $7.780 \times 10^{-5}$ & $2.520 \times 10^{-4}$ & $5.476 \times 10^{-4}$\\
			$\beta_{2}$ & $-3.492 \times 10^{-4}$ & $7.580 \times 10^{-5}$ & $-4.714 \times 10^{-4}$ & $-1.976 \times 10^{-4}$\\ 
			$\beta_{1}+\beta_{2}$ & $5.840 \times 10^{-5}$ & $8.895 \times 10^{-6}$ & $5.447 \times 10^{-5}$ & $7.619 \times 10^{-5}$\\
			\hline\hline
		\end{tabular}
	}
\end{table*}

In addition to the distance traveled in the corner area $s_c$ and entering speed $v_0$ shown in Table~\ref{table-ProbModel_input_test}, we introduced additional input parameters to the probabilistic model: preparation time $t_{p}$, positioning time $t_{q}$, breakpoint $s_{\times}$, and fatigue coefficient curve slopes $\beta_1$ and $\beta_2$. The preparation time is length of time required to prepare a patient bed for movement from CCU and the positioning time is the time taken to place a patient bed at the place of safety after its arrival. We set the mean and standard deviation of $t_{p}$ and $t_{q}$ based on the observation reported by Strating (2013)~\cite{Strating_thesis2013}, and based on Tables~\ref{table-PLRM} and~\ref{table-PLQM} for $s_{\times}$, $\beta_1$, $\beta_2$, and $\beta_1+\beta_2$.

\begin{figure}
	\centering\includegraphics[width=\columnwidth]{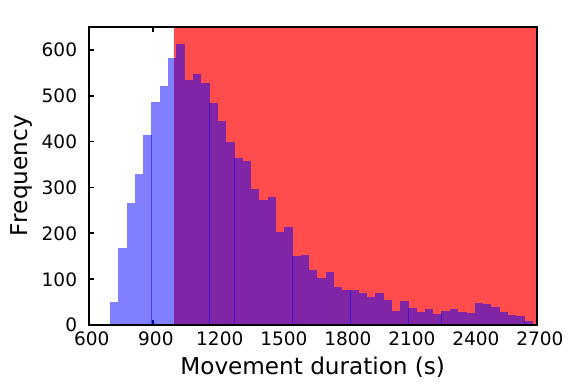}
	\caption{Numerical simulation results of movement duration for five round trips between CCU and the place of safety. The probability of finishing the movement within the reference time $T_{ref} = 999.09$~s is 23.6\%. However, 76.4\% of prediction results are longer than the reference time (indicated by the red shaded area), indicating that the reference time is not likely to be achievable. Here, the reference time $T_{ref}$ is computed with the traditional deterministic evaluation method using the mean value of input parameters.} 
	\label{fig:travel_time_case} 
\end{figure}

\begin{table}
	\normalsize                 %
	\setlength{\tabcolsep}{6pt} %
	\centering
	\caption{The probability and the corresponding available safe egress time (ASET) that a handler group can safely evacuate five bedridden patients. Note that the handler group can transport one bedridden patient to the place of safety for each round trip.}
	\label{table-EvacProbability}
	\resizebox{5.5cm}{!}{
		\begin{tabular}{cr}
			\hline\hline 
			\multirow{2}{*}{Probability} & \multicolumn{1}{c}{corresponding} \\
										 & \multicolumn{1}{c}{ASET (s)}\\
			\hline
			10\%	& 892.89 \\
			20\%	& 973.72 \\
			30\%	& 1041.17\\
			40\%	& 1109.65\\
			50\%	& 1181.31\\
			60\%	& 1266.27\\
			70\%	& 1368.30\\
			80\%	& 1506.48\\
			90\%	& 1773.08\\
			\hline\hline
		\end{tabular}
	}
\end{table}

Based on our movement duration prediction model with the probabilistic model input parameters in Table~\ref{table-ProbModel_input_case}, we performed 10000 simulations. In each simulation, the handler group makes five round trips between CCU and the place of safety. The frequency of histogram of the movement duration is shown in Fig.~\ref{fig:travel_time_case}, which has a mean of 1272.5~s and a standard deviation of 375.46~s. The estimated movement duration ranges from 718.1~s to 2699.7~s. One can observe a large spread of the numerical simulation results presented in Fig.~\ref{fig:travel_time_case} due to the right-skewed distributions of $t_c$ and $t_s$. We refer the readers to Appendix~\ref{sec:movement_duration_distribution} for further details.

By introducing a probabilistic model of movement duration prediction, we can consider the effect of uncertain factors in evacuation process, like entering speed, preparation time, positioning time, and fatigue effect coefficients. Based on the results shown in Fig.~\ref{fig:travel_time_case}, one can estimate the probability that an evacuation can be safely performed within certain amount of time. Table~\ref{table-EvacProbability} shows the probability and the corresponding available safe egress time (ASET) that a handler group can safely evacuate five bedridden patients. Note that the handler group can transport one bedridden patient to the place of safety for each round trip. For instance, the probability of successful evacuation is estimated as 90\% if ASET is 1773.08~s. In fire evacuation studies, ASET is the duration measured from the start of a fire to the onset of hazardous conditions in which people cannot be evacuated due to the fire. ASET is influenced by various factors including fire detection devices, location of fire origin, growth of fire, and ventilation paths~\cite{Cooper_FSJ1983}.

In order to quantify the effect of uncertainty in the input parameters, we compared the movement duration prediction results against the reference time which is based on the traditional deterministic evaluation method. As stated in Zhang~\textit{et al.}~\cite{Zhang_SafetySci2017}, the value of reference time was computed based on the mean value of input parameters: evacuation route length ($l = 99.3$~m), preparation time ($t_{p} = 7.96$~s), positioning time ($t_{q} = 6.25$~s), and entering speed ($v_0 = 1.07$~m/s). For a handler group making five round trips between CCU and the place of safety, the reference time $T_{ref}$ is computed as 
\noindent
\begin{equation}
\begin{split}
T_{ref} = M (t_{p}+t_{q}+t_{t}),
\end{split}
\end{equation}
\noindent
where $M$ is the number of round trips that the handler group makes and $t_t = 2l/v_{0}$ is the time required for the handler group to make a round trip between CCU and the place of safety excluding the time for preparation and positioning. As can be seen from Fig.~\ref{fig:travel_time_case}, the probability that the handlers evacuate five bedridden patients within the reference time $T_{ref} = 999.09$~s is 23.6\% while 76.4\% of prediction results are longer than the reference time. It appears that the traditional deterministic evaluation method based on the mean value of input parameters underestimates the movement duration.

\subsection{Limitations and future work}
\label{subsec:limitations}
The experiment data was collected from a small number of handlers and they are all young adults who are not professionally trained. When we set up the experiment, we were mainly interested in observing difference in male and female groups, so we did not consider a group of mixed genders. The weight of the bedridden patient was fixed and medical equipment was not connected to the patient. The fatigue effect was modeled as a function of distance traveled only, not considering other factors such as the temperature and stress level perceived by the handlers. In addition, the movement speed at which the handlers can move might be affected by the mental and physical condition of the patients, for instance, the anxiety of patients and use of intravenous fluid drips. The experiment results might be different for the mix of genders in the handler groups, weight of bedridden patient and medical equipment, and handlers’ proficiency in moving patient beds. It is also noted that the handlers in the experiment were not necessarily in a rush as they would be in the real emergency evacuation. Thus, there exists the possibility that the movement speed of handlers in the real emergency evacuation can be higher than the speed reported in this study. The presented experiment results can be generalized with larger number of handler groups having different conditions like age and proficiency.

There are noteworthy limitations in terms of the study scope. This study was performed for the case of a patient bed movement. The patient bed movement was not interrupted by other patient beds and ambulant pedestrians, thus their influence was not reflected in the presented results. In emergency evacuations, multiple groups of handlers are trying to evacuate bedridden patients in a short time and they are likely to move in the same evacuation route at the same time. In addition, some healthcare facilities do not have a dedicated evacuation route for the bedridden patients, so it is expected that a considerable number of ambulant pedestrians share the evacuation routes with the bedridden patients. Consequently, the number of interactions among patient beds and ambulant pedestrians is increasing, potentially leading to longer evacuation time. Further experimental studies need to be carried out in order to consider the interactions among bedridden patients and ambulant pedestrians and their impact on the evacuation performance of patient beds. 

Future studies can be also planned from the perspective of emergency evacuation simulations. Hunt~\textit{et al.}~\cite{Hunt_FireTech2020} pointed out that the movement of evacuation devices has not been appropriately considered in existing evacuation simulation models although the movement of such devices is critical for vulnerable patients. We are currently developing an evacuation simulation model that can incorporate our findings in patient bed movement dynamics to explicitly reflect their movement during the emergency evacuations in healthcare facilities. The evacuation simulation models can check potential conflicts with other evacuees and geometric elements such as doors and corners, and predict the total evacuation time for the scenarios in which patient beds are fleeing with other evacuees.

\section{Conclusion}
\label{conclusion}
We performed a series of controlled experiments to study the dynamics of patient beds in horizontal movement. In our experiments, we examined the change of velocity in corner turning movements and speed reductions in multiple trips between both ends of a straight corridor. In the corner turning movements, we observed that the start and end of turning movement can be identified based on the trajectory deviation. We also quantified common patterns in different speed curves by means of the normalized speed profile with rescaled time. In the straight corridor experiment, we discovered experimental relationship between the fatigue coefficient and distance traveled $s \leq 882$~m. Based on the experiment results, we developed a movement duration prediction model and then applied the model for a patient bed horizontally moving in a healthcare facility. In order to reflect uncertainty in the horizontal movement, we introduced a probability distribution to the horizontal movement parameters like entering speed, preparation and positioning time, and fatigue coefficients. According to our case study results, one can estimate the probability that an evacuation can be safely performed within certain amount of time. In addition, it is highly probable that the horizontal movement duration would be longer than the prediction results from an existing model which assumes constant movement speed. The case study results demonstrated that our model has potential in predicting emergency evacuation time of patient beds in healthcare facilities. 

As a first step to study the dynamics of patient bed movement in horizontal space, we focused on the case of a patient bed movement based on simple experiment setups. Due to the scope and setups of this study, the presented results are not applicable to the scenarios in which the movement of the patient bed is interrupted by other patient beds and ambulant pedestrians. In addition, the readers should bear in mind that the experiment data was collected from a small number of handlers and the bedridden patient was transported without medical eqipment. To generalize the findings of this study, the presented experiment should be replicated with larger number of handlers and different conditions of bedridden patients. Further experimental studies need to be carried out with more bedridden patients and ambulant pedestrians to quantify the impact of interactions with them on the evacuation time. Another interesting extension of the presented study can be planned from the perspective of emergency evacuation simulations to predict the evacuation time for various scenarios.

\section*{Acknowledgements}
This research is supported by National Research Foundation (NRF) Singapore, GOVTECH under its Virtual Singapore program Grant No. NRF2017VSG-AT3DCM001-031. The experiment was organized with the help of Ms. Yogeswary Pasupathi (Research nurse at the Singapore General Hospital Emergency Department) and members of Complexity Institute, Nanyang Technological University, Singapore. We thank Mr. Joshua for his help in processing video records.

\appendix
\section{Distribution of movement duration }
\label{sec:movement_duration_distribution}
In Sections~\ref{subsec:prediction}~and~\ref{subsec:case_study}, we presented the distribution of movement duration in a corner area and in the case study, respectively (see Figs.~\ref{fig:travel_time_prediction}~and~\ref{fig:travel_time_case}). This appendix provides evidence to suggest that the spread of movement duration can be large.

According to the setup of our numerical simulation study presented in Section~\ref{subsec:case_study}, the movement duration $T_1$ for a round trip is given as a sum of preparation time $t_p$, positioning time $t_q$, and total travel time in the corner areas $t_{c} = \sum t_{c,i}$, and total travel time in the straight corridors $t_{s} = \sum t_{s,j}$:
\begin{equation}
	\label{eq:T_1} 
	\begin{split}
		T_1	&= t_p + t_q + t_{c} + t_{s}\\
			&= t_p + t_q + \sum_{i = 1}^{N_c} t_{c,i}+ \sum_{i = 1}^{N_s} t_{s,j},
	\end{split}
\end{equation}
\noindent
where $N_c = 4$ is the number of corner areas in a round trip and $N_s$ is the number of straight corridor segments in a round trip. Travel time in corner area $i$ is denoted by $t_{c,i}$ and $t_{s,j}$ for straight corridor section $j$. Parameters $t_p$ and $t_q$ were assumed to follow normal distribution. The distribution of $t_c$ can be explained by means of the ratio distribution and the distribution of $t_s$ can be approximated by the reciprocal normal distribution. 

\begin{figure}
	\centering\includegraphics[width=\columnwidth]{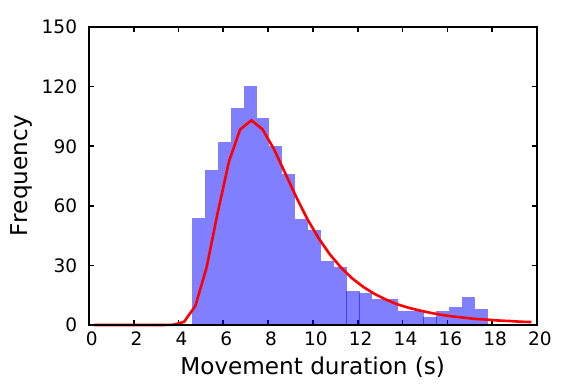}
	\caption{Frequency histogram of movement duration in a corner area estimated from numerical simulation (blue bars) and theoretically predicted by the ratio distribution (red solid line). We performed 1000 numerical simulations.} 
	\label{fig:histogram_ratioDist} 
\end{figure}

As presented in Eq.~(\ref{eq:movement_duration_final}), travel time in corner area $i$ is given as
\begin{equation}
	\label{eq:t_c,i} 
	\begin{split}
		t_{c,i} = \frac{s_c}{v_0{\hat{v}}_{avg}}.
	\end{split}
\end{equation}
\noindent
Here, the distance traveled in the corner area $s_c$ and the entering speed $v_0$ are assumed to follow the normal distribution while the average normalized speed profile ${\hat{v}}_{avg}$ is set as a fixed value based on the experimental study result. That is, $t_{c,i}$ follows a distribution of a ratio of two independent normally distributed variables $s_c$ and $v_0$. According to Curtiss~\cite{Curtiss_AMS1941} and Springer~\cite{Springer_1979}, the ratio distribution of $t_{c,i}$ can be described by means of the joint distribution of $s_c$ and $v_0$:
\begin{equation}
\label{eq:joint_sc_v0} 
\begin{split}
f(s_c, v_0) &= \frac{1}{2\pi\sigma_{s,c}\sigma_{v, 0}} \times\\
&\exp{\left\{-\frac{1}{2}\left[\left(\frac{s_c-\mu_{s,c}}{\sigma_{s,c}}\right)^2+\left(\frac{v_0-\mu_{v, 0}}{\sigma_{v, 0}}\right)^2\right]\right\}},
\end{split}
\end{equation}
\noindent
where $\mu_{s,c}=7.65$ and $\sigma_{s,c}=0.19$ are the mean and standard deviation of Gaussian distributed $s_c > 0$, and $\mu_{v, 0}=1.07$ and $\sigma_{v, 0}=0.29$ for Gaussian distributed $v_0 > 0$. The probability function of $t_{c,i}$ is given as
\begin{equation}
\label{eq:pdf_tc} 
\begin{split}
&P(t_{c,i})\\
&=\int_{-\infty}^{+\infty}{\left|v_0\right|f\left(s_c, v_0\right)dv_0}\\
&=\int_{-\infty}^{+\infty}{\left|v_0\right|f\left(v_0t_{c,i}{\hat{v}}_{avg},v_0\right)dv_0}\\
&=\frac{1}{2\pi\sigma_{s,c}\sigma_{v, 0}}\int_{0}^{+\infty}{v_0} \times\\
&{\exp{\left\{-\frac{1}{2}\left[\left(\frac{v_0t_{c,i}{\hat{v}}_{avg}-\mu_{s,c}}{\sigma_{s,c}}\right)^2+\left(\frac{v_0-\mu_{v, 0}}{\sigma_{v, 0}}\right)^2\right]\right\}}dv_0}.
\end{split}	
\end{equation}
\noindent
Based on Eq.~(\ref{eq:pdf_tc}), we theoretically predicted the frequency histogram of movement duration in a corner area and compared the result against one obtained from the numerical simulation, see Fig.~\ref{fig:histogram_ratioDist}. As can be seen from Fig.~\ref{fig:histogram_ratioDist}, the ratio distribution produced a right-skewed curve and the numerical simulation result is in good agreement with the theoretical prediction. 

\begin{figure}
	\centering\includegraphics[width=\columnwidth]{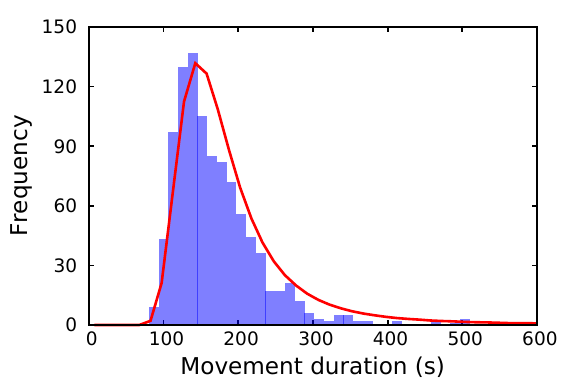}
	\caption{Frequency histogram of movement duration in the straight corridor sections (total length $s_j=168$~m) estimated from numerical simulation (blue bars) and theoretically predicted by the ratio distribution (red solid line). We performed 1000 numerical simulations.} 
	\label{fig:histogram_ReciprocalNormalDist} 
\end{figure}

As shown in Eq.~(\ref{eq:TravelTime_segment_j}), the travel time in the straight corridor section $j$ is given as
\begin{equation}
\begin{split}
t_{s,j} 
&= \frac{\Delta s_j}{v_0} \left( \frac{2}{{\hat{v}}_{j-1}+{\hat{v}}_{j}} \right)\\ 
&= \frac{2\Delta s_j}{v_{j-1}+v_{j}}.
\end{split}	
\end{equation}
\noindent
Here, ${\Delta s_j}$ is a fixed length of straight corridor section $j$, and the speed at the beginning and end points of the straight corridor section are denoted by $v_{j-1}$ and $v_j$, respectively. It is assumed that the value of $v_{j-1}$ follows the normal distribution and the value of $v_j$ is determined based on $v_{j-1}$ and the fatigue effect. It can be inferred from Section~\ref{subsec:fatigue} that the fatigue effect becomes considerable when the handler group travels several hundred meters. In a round trip, the total length of straight corridor sections is $s_j=\sum \Delta s_j=168$~m, thus it may be reasonable to suppose that the difference between $v_{j-1}$ and $v_j$ is not significant. For simplicity, we approximate $v_j$ to $v_{j-1}$ and then simplify the equation of $t_{s,j}$ as a function of the total length of straight corridor sections $s_j$ and the speed at the beginning point $v_{j-1}$:
\begin{equation}
\label{eq:t_s_j} 
\begin{split}
t_{s,j} \approx \frac{s_j}{v_{j-1}},
\end{split}	
\end{equation}
\noindent
indicating that the distribution of $t_{s,j}$ is can be described by a reciprocal normal distribution. According to Gurarie~\textit{et~al.}~\cite{Gurarie_Ecology2009}, the probability distribution function of $t_{s,j}$ is given as
\begin{equation}
\label{eq:pdf_ts} 
\begin{split}
P(t_{s,j})= \frac{s_j}{\sqrt{2\pi}\sigma_{v,j-1}t_{s,j}} \exp \left[-\frac{1}{2}\left(\frac{s_j-\mu_{v,\ j-1}t_{s,j}}{\sigma_{v,j-1}t_{s,j}}\right)^2\right],
\end{split}	
\end{equation}
\noindent
where $\mu_{v,\ j-1}=1.07$ and $\sigma_{v,j-1}=0.29$ are mean and standard deviation of $v_{j-1}$, respectively. Based on Eq.~(\ref{eq:pdf_ts}), we theoretically predicted the frequency histogram of movement duration in the straight corridor and compared the result against one obtained from the numerical simulation, see Fig.~\ref{fig:histogram_ReciprocalNormalDist}. As can be seen from Fig.~\ref{fig:histogram_ReciprocalNormalDist}, the reciprocal normal distribution generated a right-skewed curve that is consistent with the numerical simulation result.

\begin{table}
	\normalsize                 %
	\setlength{\tabcolsep}{4pt} %
	\centering
	\caption{Mode values of parameters}
	\label{table-mode-values}
	\resizebox{8cm}{!}{
		\begin{tabular}{ccl}
			\hline\hline 
			parameters & mode & note\\
			\hline
			$t_p$	& 7.96	& Normal distribution\\
			$t_q$	& 6.25	& Normal distribution\\
			$t_c$	& 29.6	& Ratio distribution\\
			$t_s$	& 138.7 & Reciprocal normal distribution\\
			\hline\hline
		\end{tabular}
	}
	\vspace{-0.5cm}
\end{table}

We computed an estimate of the movement duration for a round trip by summing the mode values of $t_p$, $t_q$, $t_c$, and $t_s$ shown in Table~\ref{table-mode-values}, i.e., $T_{1,est} = 7.96+6.25+29.6+138.7 = 182.51$~s. One can notice that the total travel time in the corner and straight corridor areas (i.e., $t_c$ and $t_s$) take a significant portion of movement duration $T_{1,est}$. Due to the right-skewed distributions of $t_c$ and $t_s$, the spread of the simulated results presented in Figs.~\ref{fig:travel_time_prediction}~and~\ref{fig:travel_time_case} is large. Given that the movement duration is inversely related to the entering speed, a small decrease in the entering speed might yield a large increase in the movement duration, extending the boundary of the histogram graph rightward.

\newpage
\bibliographystyle{model1-num-names}

\end{document}